\documentclass[epj]{svjour}
\usepackage{textcomp}
\usepackage{amssymb} 
\usepackage{graphics} 
\begin{document}  
\title{Non-classical
behavior in multimode and disordered transverse structures in OPO}
\subtitle{Use of the Q representation} 
\author{Roberta
Zambrini\inst{1} \and Stephen M. Barnett\inst{2}  \and Pere Colet \inst{1}
\and Maxi San Miguel \inst{1}}  

\offprints{roberta@imedea.uib.es} 

\institute{Instituto Mediterr\'aneo de Estudios Avanzados, IMEDEA
(CSIC-UIB), Campus Universitat Illes Balears, E-07071 Palma de Mallorca,
Spain.  http://www.imedea.uib.es/PhysDept/  \and Department of Physics,
University of Strathclyde,  Glasgow G4 0NG, United Kingdom} 
\date{Received: date / Revised version: date}  

\abstract{ We employ the Q
representation to study the non-classical correlations that are present 
from below to above-threshold in the degenerate optical parametric 
oscillator.  Our study
shows  that such correlations are present just above threshold, in the
regime in which stripe patterns are formed, but that they also persist
further above threshold in the presence of  spatially disordered
structures. 
\PACS{       {42.50.Lc}{Quantum fluctuations, quantum noise, 
and quantum jumps}  
\and       {42.50.Dv}{Nonclassical field states; squeezed, 
antibunched, and sub-Poissonian states;  
operational definitions of the phase of the field; phase measurements} 
\and  {42.65.Sf}   {Dynamics of nonlinear optical systems; optical 
instabilities, optical chaos  
 and complexity, and optical spatio-temporal dynamics}  }}

\authorrunning{Zambrini} 
\titlerunning{DOPO in Q representation} 

\maketitle  
\section{Introduction} \label{intro}   
Quantum correlations in
transverse spatial patterns of optical systems have been much studied in
recent years.  This is because  of their fundamental relevance as a
macroscopic quantum phenomena in spatially extended systems, and also
because of possible applications for quantum image processing
\cite{OPF,kolobov,optexpspecissue,qimaging}. In this context a prototype
system studied is the Optical Parametric Oscillator (OPO). 
Pattern formation in OPO was predicted long ago 
\cite{oppo}, and there are recent experimental observations of spatial
mode interaction and pattern formation \cite{fabre}.  

Quantum correlations
in the OPO are often understood in terms of the basic  process of emission
of twin signal photons  in parametric down conversion of a pump photon 
\cite{graham,tb}.  Below, but close to the threshold for parametric
oscillation, any critical transverse mode $\vec{k_c}$ is weakly damped.
The ring with radius $|\vec{k_c}|$ seen in the far field is a noisy
precursor \cite{wiesenfeld} of the wave-number to be selected above threshold.
Spatial correlations have been studied in this
regime  showing interesting non-classical behavior \cite{marzoli97}.
This is referred in the literature as a quantum image
\cite{quantumimages}.
In particular, the intensity difference between any two opposite modes 
with radius $|\vec{k_c}|$ in the far field is subpoissonian. 
This is one manifestation of quantum
entanglement of the  modes \cite{notetb}. From a technical point of view,
correlations in the regime below threshold can be calculated analytically
within linear approximations and using the Wigner representation
\cite{OPOale}. Above, but relatively close to threshold, a stripe pattern
appears (Sect. \ref{Class}) due to the interference of two selected signal
modes of opposite critical transverse wavenumber \cite{OPOabove}. Again,
this result can be understood as a macroscopic consequence of the
microscopic process of twin photon emission.  Close to threshold this
phenomenon is studied with models which consider only a small number
of modes; only the pump mode and the critical ones are considered.  The
calculation is based on a linearization approximation and the use of the
Wigner representation. At threshold or extremely close to threshold,
critical fluctuations would invalidate a linearized calculation, but a few
mode approximation, such as those considered in \cite{critical}, could be
appropriate.  

Quite a different situation exists when multimode nonlinear
dynamics have to be considered. Some results have been obtained in the
context of $multimode$ interaction in linear approximations in other
systems \cite{zambrini-kerr,gattihexagonos,damia}.  However, a full
multimode nonlinear description is needed well above threshold where the
emission of the critical modes necessarily stimulates spatial harmonics in
both the pump and the signal, and undamped modes associated to spatial
broken symmetries play an important role. This is also the case in the
convective regime caused by walk-off, where a noise sustained pattern
appears with a broad  (multimode) spectrum and amplified nonlinear
fluctuations \cite{marco,zambrini}. In these situations two relevant
questions have to be addressed. First, from the conceptual point of view
one might ask if the quantum correlations are degraded by the subsidiary
nonlinear processes of multimode competition. In particular, can the twin
beams that appear close to threshold be incoherently depleted by other
processes?  If, however, quantum effects persist, then the simple
explanation in terms of emission of pairs of twin photons needs to be
revised. Secondly, and from a technical point of view, new calculation
techniques or approximations that go beyond linearized approximations need
to be developed.  

We have previously examined these questions using a
time-dependent parametric approximation \cite{zambrini,twin}. We concluded
that quantum correlations were strongly suppressed in the noise-dominated
convective regime \cite{zambrini}. However, in the regime of absolute
instability we found that non-classical correlations 
between the $\pm k_c$ signal beams
persists in spite of nonlinear interactions with the pump and with higher
order harmonics of the signal \cite{twin}. This is still true in cases in
which the mean intensity of the two critical signal beams is different
because of walk-off \cite{twin}. Two limitations of our time dependent
parametric approximation, which is well suited in the convective regime,
are that fluctuations in the pump are neglected and that its validity is
restricted to values relatively close to threshold.  

As an alternative
approach we propose, in this paper, the use of the Q-representation and
its associated nonlinear Langevin equations for c-number complex fields.
With this method we can study quantum spatial correlations in the DOPO
without any linearization, few mode approximation or time-dependent
parametric  approximation, for values of the pump less than twice its
threshold value. In particular we can reach high pump values for which the
stationary classical solution for the signal field is not a stripe
pattern, but rather a homogeneous state with one of two preferred phases.
Spatial coexistence of domains of these two homogeneous solutions
separated by domain walls gives rise to  spatially disordered patterns
(seemingly chaotic).  These have a broad spectrum in the far field for
both the pump and the signal field. This structure cannot be described,
not even as a first approximation, as the interference of two twin beams
of opposite transverse critical wavenumber.  This is because of other
cascading processes coupling many different modes and frequencies.
Nevertheless, we find that any $\pm k$ pair of modes in the far field that
lie within the active range of the broad spectrum show non-classical 
correlations. This
seems to indicate that the basic process of parametric down conversion is
behind correlations found even in these very complicated spatial
structures.  

The paper is organized as follows.  The quantum formulation
of the problem is presented in Sect.
\ref{Q.f.}, where phase space methods are 
briefly compared (Sect. \ref{M.e.}) and Langevin equations for the
Q representation are introduced (Sect. \ref{Q.repr}). 
In Sect. \ref{Class}  we briefly review relevant classical
results concerning the instabilities to pattern formation and to
homogeneous solutions well above threshold. Our general aim is to study
the quantum properties of the correlations in regimes that have not been
previously studied.  These include the critical point in presence of a
multimode interaction (Sect. \ref{below}), and the  above threshold region
in presence of complex patterns (Sect. \ref{above}). We find signatures of
entanglement of the beams in the presence of stripes influenced by the
presence of higher harmonics (Sect. \ref{stripe}). We show that
such entanglement is not only associated with a stripe pattern, but it is
present also in spatially disordered structures (Sect.
\ref{disordered}).  Finally   Sect. \ref{conclusions} is devoted to
concluding remarks. 

\section{Quantum formulation of DOPO dynamics} 
\label{Q.f.}  

To describe the intracavity dynamics in a  
DOPO we introduce the boson
spatial modes $\hat A_0(\vec x,t)$ and $\hat A_1(\vec x,t)$, respectively
at the pump frequency $2\omega$, and signal frequency $\omega$, and
satisfying standard equal-time commutation relations \cite{OPOale}
\begin{equation} 
\left[\hat A_i(\vec x, t), \hat A_j^\dagger(\vec x',
t)\right]=\delta_{ij}\delta(\vec x- \vec x') \, , ~~~i,j=0,1
\label{equaltimecommutator} 
\end{equation} 
Here $\vec x$ denotes the
transverse coordinate(s). A Hamiltonian operator describes the interaction
between these modes in the non-linear medium. The intracavity fields
constitute an open device \cite{carmichael,steve}, modeled within a
statistical approach in the Schr\"{o}dinger picture by a Master equation.
In Sect.\ref{M.e.} we review the Master equation of a DOPO. We then report
on the possible phase-space descriptions, introducing the Q-representation
and the associated Langevin equations (Sect.\ref{Q.repr}). 

\subsection{Master equation and phase space descriptions} 
\label{M.e.} 

The intracavity dynamics of our open system is described by a Master
equation for the reduced density operator $\hat{\rho}$
\cite{carmichael,steve}: 
\begin{equation} 
\label{master}
\frac{\partial\hat \rho}{\partial t}=\frac{1}{i\hbar}[\hat H, \hat
\rho]+\hat \Lambda\hat\rho \; . \end{equation} 
We consider a plane
one-sided cavity, hence the Liouvillian accounting for dissipation through
the partially reflecting mirror is given by 
\begin{eqnarray}\nonumber \hat
\Lambda\hat \rho = \sum_{i=0,1}\gamma_i\int d^2\vec x \left\{[\hat
A_i(\vec x),\hat  \rho \hat A_i^\dagger(\vec x)]+[\hat A_i(\vec x)\hat
\rho,\hat  A_i^\dagger(\vec x)] \right\} \label{liouv} \; . 
\end{eqnarray}
The  Hamiltonian  o\-pe\-ra\-tor, ex\-pres\-sed as a function of fields operators  
$\hat A_0(\vec x,t)$ and $\hat A_1(\vec x,t)$, is: 
\begin{equation} \hat
H=\hat H_0+\hat H_{int} +\hat H_{ext} \end{equation} where
\cite{stevenote} \begin{equation}\hat  H_0=\hbar \int{d^2 \vec x
\sum_{i=0,1} \left[\gamma_i \hat {A}_i^\dag(\vec
x)(\Delta_i-a_i\nabla^2)\hat {A}_i(\vec x) \right]} \, 
\end{equation}
describes free propagation of fields in the cavity, 
\begin{equation} \hat
H_{ext}=i \hbar \int{d^2\vec x E\left[\hat {A}_0^\dag(\vec x)- \hat
{A}_0(\vec x)\right]} 
\end{equation} 
is due to the interaction with the
external pump $E$, which we choose  to be real, and 
\begin{equation} \hat
H_{int}=i\hbar \frac{g}{2}\int {d^2 \vec x \left[\hat {A}_0(\vec x)
\hat{A}_1^{\dag 2}(\vec x)- {\hat {A}_0}^\dag (\vec x){\hat {A}_1}^2(\vec
x)\right]} 
\end{equation} 
is the interaction term between first and second
harmonic.  

Density operators in state space
can be mapped to quasi-probability distribution densities on phase space.
These can be used to calculate ensemble averages of operators in defined
orderings \cite{schleich,gardiner}. Using this ``quantum-to-classical"
correspondence, Eq. (\ref{master}) can be converted into an equation of
motion for a quasi-probability distribution in the phase-space of
radiation fields $\alpha_i(\vec x)$, associated with the operators $\hat
{A}_i(\vec x)$.  

The presence of non-linearities leads to a functional
differential equation for the quasi-probability that is not of the
Fokker-Planck  type, leading to difficulties in obtaining solutions
\cite{gardiner.stoc}. The offending term for the system of interest is
$\hat{H}_{int}$, which gives a functional term of the form
\begin{eqnarray}\label{nl_term} &&[\hat {A}_0(\vec x)\hat
A_1^{\dag^2}(\vec x)- h.c.,\hat \rho]\Longleftrightarrow
\left(s\alpha_0\frac{\delta^2}{\delta \alpha_{1}^2}+\right.\\
&&\left.\frac{1-s^2}{4}\frac{\delta^3}{\delta \alpha_{1}^2 \delta
\alpha_{0}^*}+\frac{\delta}{\delta \alpha_{0}}\alpha_1^2-
2\alpha_{0}\alpha_1^*\frac{\delta}{\delta \alpha_{1}}+c.c.\right)W_s.
\nonumber \end{eqnarray} Here $s$ denotes the operator ordering
selected. 
We see that third order derivatives appear in the temporal evolution of
the Wigner representation ($s=0$). The approximate equations obtained
simply dropping these  third order terms constitute the basis of
{\it stochastic
electrodynamics} \cite{stoc.el}.  This approach works well in {\it linear}
regimes, in which the Wigner  distribution satisfies a genuine
Fokker-Planck equation. In particular, in the DOPO below the threshold of
signal generation the intensity of the signal is of the order of the
quantum noise, while the pump has a macroscopic mean value, so that its
fluctuations can be neglected.  With the assumption of a classical
undepleted pump, the Hamiltonian that describes the quantum dynamics of
the signal is quadratic and the Langevin equation \cite{gardiner.stoc} --
equivalent to the Fokker Planck equation for  Wigner representation -- can
be analytically solved \cite{OPOale}. Recent investigations have shown the
limits of this stochastic electrodynamics in reproducing  quantum
higher-order moments  \cite{3ord.mom}. 
The same type of approximation is
generally possible above threshold, linearizing around a pattern solution
\cite{zambrini-kerr}. Due to the in-homogeneity of the reference state,
moreover, only semi-analytical or
numerical simulations can be provided. Above threshold
great care has to be taken if the system is translational invariant
(flat mirrors and homogeneous  transversal  pump profile). The
pattern solution breaks the translational symmetry and therefore 
there is a
Goldstone mode which is neutrally stable  \cite{zambrini-kerr}. Noise
excites this mode, giving diffusion of the phase which fixes the
position of the pattern \cite{zambrini-kerr}. Moments involving such big
fluctuations cannot be correctly described within a linearized treatment
in the fields amplitudes.  In particular, such an approach leads to
unphysically divergent quadrature correlations, although correct results
can be obtained for the intensity correlations \cite{damia}.  

Another
well-known quasi-probability is the P${_+}$ representation \cite{P+},
consisting in the extension of the normal ordered P representation over a
doubled phase space \cite{doubling}.  The P representation ($s=1$ in Eq. (\ref{nl_term}))
suffers of negative diffusion in  problems of interest, but the P${_+}$
generally gives good results, with the advantage of the possibility to
obtain immediately also the moments outside the cavity. However, in some
systems  this doubling phase-space technique has shown divergent
trajectories, as reviewed in Ref. \cite{P+review}.  In particular,  there
are regimes in extended systems --like the convective regime
\cite{zambrini}-- in which the
presence of large fluctuations  around the unstable reference state would
result in diverging trajectories in the P${_+}$ and
alternative methods are needed.

In this paper we
employ the Q representation corresponding to anti-normal ordering of field
operators. The most important property of this representation is that it
satisfies the requirements for a true probability distribution. In fact
the Q-representation may be defined as the diagonal matrix elements of the
density operator in the space of coherent states 
\begin{eqnarray}
\label{Qdef}
Q(\alpha_0,\alpha_1)=\frac{1}{\pi}<\alpha_0,\alpha_1|\hat{\rho}|
\alpha_0,\alpha_1> 
\end{eqnarray} 
and so is both positive and bounded
\cite{gardiner}. Due to the over completeness of the coherent states
ensemble, the definition (\ref{Qdef}) uniquely determines the density
operator $\hat{\rho}$.  Physically this representation, resulting from a
Gaussian convolution of the Wigner representation, corresponds to
simultaneous measurements of orthogonal quadratures, as limited by the
Heisenberg principle, in a  eight-port homodyne detector \cite{Q-4ports}.
From Eq. (\ref{nl_term}) (with $s=-1$) we observe that the
Q-representation suffers of negative diffusion.  Unlike the Wigner
function, however, the Q function is always positive and well-behaved. The
possibility to obtain a positive solution in presence of a negative
diffusion \cite{pawula} lies in the presence of a restricted ensemble of
initial conditions, that cannot be arbitrarily narrow. In other words, not
all mathematical forms for the Q function correspond to physical states.
The evolution of a physical state - corresponding to an hermitian density
operator - in the Q representation will always be positive
\cite{gardiner,risken}. We should note that a Q representation with a
doubled phase-space has been proposed in order to deal with negative
diffusion \cite{Yuen}.  This has been shown to give good results in some
non-linear quantum systems  \cite{ro+steve}.  

In  the next section we
investigate  the possibility to use the Q-representation for devices
consisting on a cavity filled with a $\chi^2$ medium, as in the OPO and
Second Harmonic Generation (SHG).   

\subsection{Langevin equations in Q-representation} 
\label{Q.repr}   

The evolution equation of the
functional Q for our model (described above) is: \begin{eqnarray}\nonumber
&&\frac{\partial Q(\alpha_0,\alpha_1)}{\partial t}=\int d^2 \vec x \left\{
-\left(\frac{\delta}{\delta \alpha_0(\vec x)} V_0+\frac{\delta}{\delta
\alpha_1(\vec x)}V_1 +c.c.\right)\right.\\\nonumber &&\left.+ \int d^2
\vec x'\left[ 2\gamma_0\frac{\delta^2} {\delta \alpha_0(\vec
x)\delta\alpha_0^* (\vec x')}+ 2\gamma_1\frac{\delta^2}{\delta
\alpha_1(\vec x)\delta\alpha_1^* (\vec x')}+\right.\right.\\
&&\left.\left. \frac{1}{2}\left(-g\alpha_0\frac{\delta^2}{\delta
\alpha_1(\vec x)\delta\alpha_1(\vec x')}
+c.c\right)\right]\right\}Q(\alpha_0,\alpha_1), \label{Q.eq}
\end{eqnarray} where the drift terms are \begin{eqnarray}\nonumber
V_0&=&-\gamma_0[(1+i\Delta_0)-ia_0\nabla^2] \alpha_0(\vec x,t)-\frac{g}{2}
\alpha_1^2(\vec x,t)+E\\\nonumber
V_1&=&-\gamma_1[(1+i\Delta_1)-ia_1\nabla^2] \alpha_1(\vec x,t)+g
\alpha_0(\vec x,t)\alpha_1^*(\vec x,t). \end{eqnarray} 
If the
diffusion term is positive then our evolution  equation is a 
bona-fide Fokker-Planck
equation. In the other case this equation doesn't describe an ordinary
diffusion process. For equation (\ref{Q.eq})  the diffusion term is
positive if 
\begin{equation} \label{d.positive} 
|\alpha_0(\vec
x,t)|<\frac{2\gamma_1}{g} . 
\end{equation} 
The modulus of the stationary field at
threshold takes the value $|A_0^{thr}|={\gamma_1}/{g}$ (see Sect.
\ref{Class}). This means that the condition (\ref{d.positive}) corresponds
to pump trajectories taking values that are less than twice the threshold
value. Staying in a  region far from the limit (\ref{d.positive}) -- we
are considering $A_0^{st} \leq 1.5A_0^{thr}$ -- an extremely large
fluctuation in a trajectory would be necessary in order to lose the
positiveness of the diffusion. Clearly these trajectories have a
negligible probability to appear, and never appeared in our simulations.
For these reasons, the approximation we propose is to study Langevin
equations related to the Fokker-Planck equation given by (\ref{Q.eq})
$and$ (\ref{d.positive}), neglecting any  trajectories that would make
negative the diffusion term.  Clearly the condition (\ref{d.positive})
does not depend on the frequency at which the system is pumped.  For this
reason the method is suitable, and has been already successfully used, to
describe non-linear fluctuations in stripe patterns in SHG in a regime of
pump values limited by Eq. (\ref{d.positive}) \cite{morten}.  

From
Eqs.(\ref{Q.eq}-\ref{d.positive}), with the scaling 
(D is the transversal dimensionality of the system)
\begin{eqnarray} \label{scaling} 
&&\gamma_0=\gamma_1=\gamma~,~~~~
a_0=a_1/2=a, \\ 
&&t'=\gamma t~,~~~~\vec x'=\frac{\vec x}{\sqrt a}, \nonumber\\
&&A_i'=\frac{g}{\gamma}A_i~,~~~~ E'=\frac{g}{\gamma^2}E~,~~~~
\epsilon_i'=\frac{g}{\gamma^{3/2}  a^{D/4}}\epsilon_i, \nonumber
\end{eqnarray} 
we obtain the equations: 
\begin{eqnarray}  \partial_t
\alpha_0(\vec x,t)&=& - \left[(1+i\Delta_0)-i\nabla^2 \right]\alpha_0(\vec
x,t)+E-\nonumber\\ \label{Eq:lang0} &&\frac{1}{2}\alpha_1^2(\vec x,t)
+\sqrt{\frac{2}{a}}\frac{g}{\gamma}\xi_0(\vec x,t)\\ \nonumber \partial_t
\alpha_1(\vec x,t)&=& - \left[(1+i\Delta_1)-2i\nabla^2
\right]\alpha_1(\vec x,t)+\\ \label{Eq:lang1} &&\alpha_0(\vec x,t)
\alpha_1^*(\vec x,t)+\sqrt{\frac{2}{ a}}\frac{g}{\gamma}\xi_1(\vec x,t).
\end{eqnarray} 
The condition (\ref{d.positive}) in the new variables is
\begin{equation} \label{d.positive.sc} 
|\alpha_0(\vec x,t)|<2. 
\end{equation} 
We solve these Langevin equations by numerical simulation,
neglecting any trajectories that  do not satisfy the condition
(\ref{d.positive.sc}), should these occur. 
$\xi_0$ is a white Gaussian noise with non-vanishing moment:
\begin{eqnarray}
\langle \xi_0(\vec x,t)\xi_0^*(\vec x',t') \rangle&=& \delta(\vec x-\vec
x')(t-t')
\end{eqnarray} 
The signal noise $\xi_1$ results to be  {\it phase sensitive}, 
due to the presence of diagonal
terms in the diffusion matrix of Eq. (\ref{Q.eq}).
Moreover this noise is multiplicative,  depending on the value of
the pump field.  However due  to the form of
$\hat H$ (quadratic in $\hat A_1$ and linear in $\hat A_0$), these
equations have the same formal expression in the Ito or Stratonovich 
interpretations \cite{gardiner.stoc}. 

The phase sensitive multiplicative noise $\xi_1(x,t)$ can be written as
\begin{eqnarray} \label{x1}
\xi_1(x,t)&=& \left[\frac{-\alpha_{0I}(x,t)}{2\sqrt{2+
\alpha_{0R}(x,t)}}+
{i\over 2}\sqrt{2+\alpha_{0R}(x,t)}\right] \phi(x,t)
\nonumber\\
&+&\sqrt{\frac{1-{ |\alpha_{0}(x,t)|^2\over 4}} {2+\alpha_{0R}(x,t)}    
}\psi(x,t) 
\end{eqnarray} 
with $\alpha_{0}=\alpha_{0R}+i\alpha_{0I}$ and $\phi,\psi$ 
uncorrelated real
white noises in space and time, with variances one. 
Actually, the diffusion matrix of a Fokker-Plank equation
fix only three of the four degrees of freedom in the choice
of the real and imaginary parts of the noise term  $\xi_1$; this depends
on the multiplicity of Langevin processes associated with the same
Fokker-Planck equation \cite{gardiner.stoc}. Hence  Eq. (\ref{x1})
corresponds to a particular choice among  several possible
representations \cite{obs}.   

In the next sections we present the numerical results
obtained by simulating Eq. (\ref{Eq:lang0}) and Eq. (\ref{Eq:lang1}) with
the same integration method of Ref. \cite{zambrini-kerr,morten}. In
particular we consider one transversal dimension ($D=1$) and parameters:
\begin{eqnarray}
\Delta_0=0,~~~~~\Delta_1=-0.18,~~~~~\frac{g}{\sqrt{a}\gamma}=10^{-4}
\end{eqnarray} with a system size of four critical wavelengths
$L=4\lambda_c$, with $\lambda_c=\frac{2\pi}{k_c}$ (see Sect.
\ref{Class}).  

Eqs. (\ref{Eq:lang0}-\ref{Eq:lang1}) are
suitable for the study of quantum fluctuations in different regimes  --
discussed in Sect.  \ref{Class} -- from the linear regime below
threshold to the multimode regimes quite above threshold. In Sect.
\ref{below} we present calculations of quantum correlations below
threshold and at the critical point, comparing with analytical results in
the linear approximation. In Sect. \ref{above} we consider
the above threshold regime, in the situation of regular stripe pattern
formation
and also when disordered structures are formed  (see Fig. \ref{fig:1}).

\section{Pattern formation in OPO} 
\label{Class}  

We first
review the patterns that arise in the OPO in the classical limit. The
phase matched DOPO, where both pump $A_0$ and signal $A_1$ fields are
resonated, is described classically by equations identical to  Eqs.
(\ref{Eq:lang0}) and (\ref{Eq:lang1})
but neglecting the noise terms. The classical equations have a
trivial homogeneous solution 
\begin{equation} \label{zero_hom_sol}
A_1^{st}=0, \, \, A_0^{st}={E \over 1+i\Delta_0}. 
\end{equation} 
A linear
stability analysis around this solution gives the following dispersion
relation for the growth of the signal field perturbations with wave vector
$k$ (the pump field $A_0$ is always stable)
\cite{oppo} 
\begin{equation}
\lambda_1(\vec k) = -1 \pm \sqrt{|A_0^{st}|^2-(\Delta_1+2k^2)}.
\label{dispersionrelation} 
\end{equation} 
For negative signal detunings,
the zero homogeneous solution becomes unstable at $E = E_{c} =
\sqrt{1+\Delta_0^2}$. The perturbations with maximum growth rate are those
with wave number $|k_c|=\sqrt{-\Delta_1/2}$, and a pattern with this wave
number is formed at threshold \cite{oppo}. For positive signal detunings the zero
homogeneous solution is stable for $E < E_{c} =
\sqrt{(1+\Delta_0^2)(1+\Delta_1^2)}$. In this case the instability takes
place at zero wave number leading to a non zero homogeneous solution
\cite{trillo-pettiaux-Drummond} 
\begin{eqnarray} \label{nonzero_hom_sol} 
A_1^{st} &=& \pm\sqrt{E
|A_1^{st}|^2 \over (1+i\Delta_0)(1+i\Delta_1)+|A_1^{st}|^2} \\ A_0^{st}
&=& {E (1+i\Delta_1) \over (1+i\Delta_0)(1+i\Delta_1)+|A_1^{st}|^2} \, .
\end{eqnarray} 
There are two equivalent solutions for the signal field
with a $\pi$ phase difference.  

In this paper we will
consider the case of zero pump detuning ($\Delta_0=0$) and negative signal
detuning ($\Delta_1 < 0$),  in which stripe pattern
arises at threshold ($E_{c}=1$). Increasing further
the pump  the nonzero homogeneous solutions
(\ref{nonzero_hom_sol}) become stable. In fact, we observe they are
 stable for pump values around $E=1.1$.  
This is  lower than the value at which the
stripe pattern becomes linearly unstable \cite{etrich,gomila}.  
Numerical studies in this
regime  find multistability between the stripe pattern, the two nonzero
homogeneous solutions and several irregular spatially modulated solutions.
The latter are formed by fronts with oscillatory tails connecting the two
equivalent homogeneous solutions. In systems with two spatial dimensions,
there are also coexisting labyrinthine patterns. In this case ($D=2$) both
the irregular spatially modulated solutions and the labyrinthine patterns
cease to exist at the modulational instability of a flat front connecting
the two homogeneous solutions \cite{gomila,gomilaprl}. Here we  consider
systems with only one spatial dimension for which such an instability does
not exist. This means that the regime of multistability in parameter space
is much larger. The irregular spatially modulated solutions found in this
system are an example of frozen chaos as described in Ref. \cite{Coullet}. In
that case the interaction of two distant fronts can be described by a
potential with several wells which become progressively deeper as the
distances between the fronts decreases. The OPO cannot be described in
terms of such potentials, but numerical studies reveal equilibrium
distances whenever the maxima (or the minima) of the local oscillations of
the front overlap with each other \cite{etrich,gomila}.  

\section{Quantum correlations below and at threshold} 
\label{below}  

The spatiotemporal dynamics of the signal field is shown in
Fig. \ref{fig:1}  for two relevant values of the pump, below but near to 
threshold (quantum images regime), $E=0.999$ and  at threshold
($E=1$). The far field (FF) shows strong fluctuations dominated by
the critical wave-vector: in Sects. (\ref{quadrature}-\ref{intensity}) we
discuss the  quadratures and intensity quantum correlations of these
modes.   

\subsection{Quadrature correlations}  
\label{quadrature}  

The direction in which quadrature squeezing 
appears is determined by the
eigenfunction $V_\pm( k,- k)$ of the linear problem $\partial_t V_\pm( k,-
k)= \lambda_\pm( k) V_\pm( k,- k)$, as reported in Ref. \cite{zambrini}:
\begin{eqnarray} 
\label{inst_dir} 
&&V_\pm( k,- k)=e^{i\Phi_\pm}\delta A_1(
k)\pm\delta A_1^*(- k)\\ &&e^{i\Phi_\pm( k)}= \mp\frac{i\Delta_1+2i k^2\mp
\sqrt{|A_0^{st}|^2-(\Delta_1+2  k^2)^2}} {A_0^{st}}. \nonumber
\end{eqnarray} 
The solution $V_+( k,- k)$ gives the direction of
amplification of fluctuations, while fluctuations are damped for $V_-( k,-
k)$, giving rise to quadrature squeezing. In particular, for the critical
wave-vector $ k_c$ and for our choice of parameter  (real $A_0^{st}$) we
obtain  $V_\pm(k_c,- k_c)=\delta A_1( k_c)\pm \delta A_1^*(- k_c)$.
Therefore, the largest squeezing at threshold will be in the difference of
real parts and the sum of imaginary parts of the
field for wave-numbers $ k_c$ and $- k_c$. 

We define the real quadrature operator: 
\begin{eqnarray} \hat X(k)&=& \hat
{A}_1(k) + \hat {A}_1^\dag(k) \end{eqnarray} and the quadrature
superpositions \begin{eqnarray} \label{Eq:x-} \hat X_-(k)&=& \hat X(k) -
\hat X(-k) \\ \label{Eq:x+} \hat X_+(k)&=& \hat X(k) + \hat X(-k),
\end{eqnarray} 
corresponding, respectively, to damped and undamped
quantities at threshold.  

\begin{figure}
\resizebox{0.99\columnwidth}{!}{\includegraphics{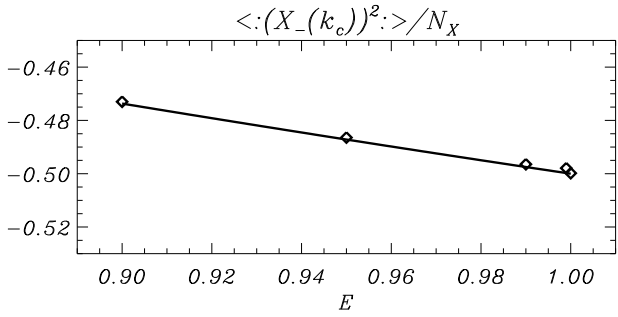}}
\caption{Normal ordered variance of the damped quadrature $\hat X_-(k_c)$
normalized to shot noise: 
diamonds are results obtained by numerical
simulation, while the continuous line corresponds to the analytical
expression Eq. (\ref{eq:squeez}). For any trajectory at given pump
intensity, we average during a time of $10^7$, integrating with a time
discretization of $10^{-3}$ (with time scaled as in  Eq. (\ref{scaling})).   }
\label{fig:2}
\end{figure}

\begin{figure}
\resizebox{0.99\columnwidth}{!}{\includegraphics{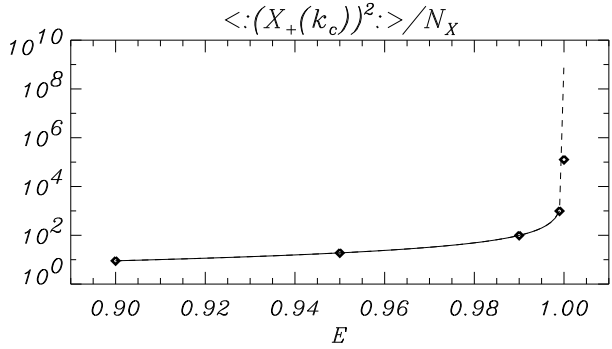}}
\caption{Variance of the undamped quadrature $\hat X_+(k_c)$: the diamonds
are results obtained with numerical simulation, while the continuous 
line corresponds
to the analytical expression Eq. (\ref{eq:unsqueez}). 
At the last point, corresponding to $E=1$, the linear
treatment gives an infinite variance (the asymptotic behavior is
represented by a dashed line), while our non-linear treatment gives the
expected saturation. } \label{fig:3} \end{figure}

Below threshold, within a linearization
approximation  \cite{OPOale}, the 
normal-ordered variances normalized to the shot noise ($\mathcal{N}_X$) 
\cite{shot-noise} are: 
\begin{eqnarray} \label{eq:squeez} \frac{<:(\hat
X_-(k_c))^2:>}{\mathcal{N}_X}&=& \frac{-E}{1+E} \\ \label{eq:unsqueez}
\frac{<:(\hat X_+(k_c))^2:>}{\mathcal{N}_X}&=&  \frac{E}{1-E}.
\end{eqnarray} 
These quantities coincide with the variances since the mean values are 
zero: $<\hat X_\pm(k)>=0$. The  normal ordering allows us to immediately
identify non-classical features associated with  squeezing such as negative
variances.  Eq. (\ref{eq:squeez}) shows an increasing degree of squeezing,
approaching the value $-0.5$ at threshold. In Fig. \ref{fig:2} theoretical
predictions and numerical results are shown to be in good agreement,
confirming the validity of Eqs. (\ref{Eq:lang0}-\ref{Eq:lang1})  below
threshold. On the other hand Eq. (\ref{eq:unsqueez}) is always positive
indicating that the the fluctuations in the direction of instability are
essentially classical and larger than those found for a coherent state. In Fig.
\ref{fig:3} we show the agreement between theoretical predictions and
numerical results for the undamped quadrature, even as close as $1$ 
\textperthousand to 
threshold. The limits of the linear treatment, discussed above, are now
evident in the divergence of Eq. (\ref{eq:unsqueez}) for $E\rightarrow
1$.  In contrast,  numerical simulation of the nonlinear Eqs.
(\ref{Eq:lang0}-\ref{Eq:lang1}) gives the expected saturation at the
critical point, at a value which depends on the noise level.  

\subsection{Intensity correlations} 
\label{intensity}  

We can find
non-classical features in the intensities  of the twin beams by evaluating
the normal-ordered variance in the difference of the two intensities:
\begin{eqnarray}\label{V} {\mathcal
V(k)}=\frac{\langle:[\delta\hat N_1(k)-\delta\hat
N_1(-k)]^2:\rangle}{\mathcal{N}_N(k)}, 
\end{eqnarray} 
normalized to the corresponding shot noise value $\mathcal{N}_N(k)$. 
This value is proportional to the sum of the intensities of the two beams
with wavevectors $\pm k$.
 Negative values of ${\mathcal V}$ indicate 
sub-Poissonian statistics for the intensity difference of the two signal
beams at $\pm k$ \cite{twin}. In a linear analytical treatment below
threshold ${\mathcal V(k)}=-0.5$, independently of the pump intensity and
of the wave-vector \cite{OPOale,twin}. In other words the normalized 
intensity correlations, Eq. (\ref{V}), do not show a
non-classical behavior which is stronger
for  the critical wave vector or at the critical
point.  This is in  contrast with the behavior of  the quadratures
correlations Eqs. (\ref{eq:squeez}-\ref{eq:unsqueez}). 
Nevertheless,  the critical conditions
are of significant interest because of  presence of higher intensities.
 
\begin{figure}
\resizebox{0.9\columnwidth}{!}{\includegraphics{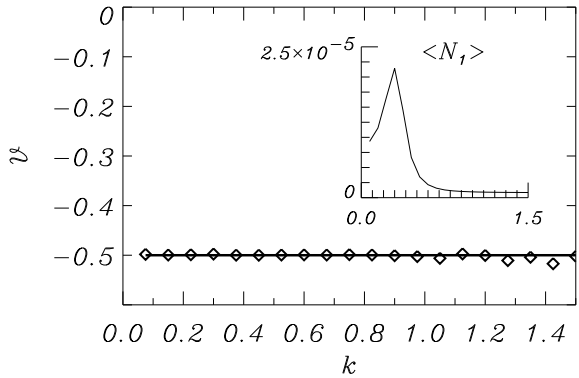}}
\caption{Analytical (continuous line) and numerical (diamonds) twin beams
correlations ${\mathcal V(k)}$ (Eq. (\ref{V})) below threshold ($E=0.99$).
The insert shows the mean intensity of the signal, with a maximum  
at $k_c=0.3$.} \label{fig:4} \end{figure}

The numerical expression of ${\mathcal V(k)}$ for different spatial modes 
($0<k\le 5k_c$) is compared, in Fig. \ref{fig:4},  with  the analytical value
$-0.5$  for pump $E=0.99$, showing good agreement.   Small
deviations  for large wavevectors $k$ can appear numerically due to the
smallness of the shot noise to which the variance is normalized;
${\mathcal{N}_N}$ is proportional to the mean intensity of  the field,
shown also in Fig. \ref{fig:4}.  

\begin{figure}
\resizebox{0.95\columnwidth}{!}{\includegraphics{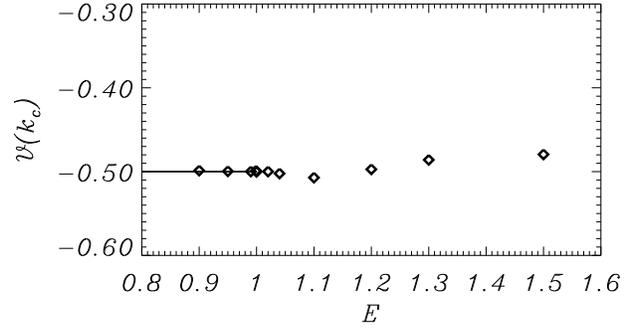}}
\caption{Variance ${\mathcal V(k_c)}$ at the critical wave-vector
($k_c=0.3$), increasing the pump intensity from below to above threshold
($E_c=1$). The diamonds are numerical results and the continuous line is
the analytical value obtained by linearizing below threshold. } 
\label{fig:5}
\end{figure}

In  Fig. \ref{fig:5} we
plot the variance  ${\mathcal V(k_c)}$ at the critical wave-vector as a
function of the pump $E$:  We obtain good agreement with analytical
predictions below threshold.   

\section{Quantum correlations above threshold} 
\label{above}   

The non-linear equations in the
Q-representation are used here to study the regime
of pattern formation above threshold.
These equations improve  the
time-dependent parametric approximation introduced in Ref. \cite{zambrini}, 
which is well suited to study the convective regime (in presence
of walk-off) or  the instability point, because the pump
fluctuations are disregarded with respect to the signal ones. 
Eqs.(\ref{Eq:lang0}) and (\ref{Eq:lang1}) are valid in a wide
region above threshold, and give a complete description of the pump
fluctuations.  

\begin{figure}
\resizebox{0.95\columnwidth}{!}{\includegraphics{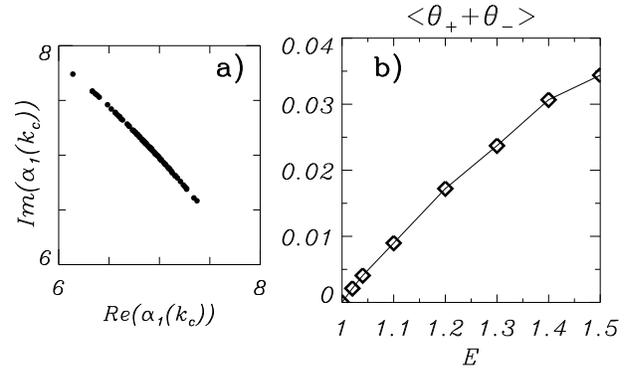}} \caption{
a) Trajectory in the phase space of $\alpha_1(k_c)$ during $10^7$ scaled
units for $E=1.02$. b) Phase sum $<\theta_++\theta_->$ (in radians) 
increasing the pump intensity.} 
\label{fig:6} \end{figure}

Slightly above the threshold  ($E=1.02$) the variance of
the quadratures superposition Eq. (\ref{Eq:x-}) becomes classical.
To understand the how the mode dynamics changes when going above threshold,
we recall that below threshold the trajectory 
$\alpha_1(k_c,t)$ occupies a circular
region centered in zero, in the phase space given by its real and
imaginary part. Above threshold the mean amplitude has a macroscopic value
increasing with the pump intensity and the 
distribution of fluctuations in phase and intensity
quadratures is rather different, as shown in Fig. \ref{fig:6}a.
A trajectory for a mode during a larger time would describe a circle in
phase space. 

The origin of these
large fluctuations in the phase qua\-dra\-tu\-re is well known in the theory of
single transverse-mode non-degenerate parametric oscillators (NDOPO)
\cite{Reid-Drummond}. Due to the diffusion of the difference of the
signal and idler phases, the above threshold solution is not stable,
and ``cannot be analyzed correctly by the assumption of small fluctuations
and methods of linearization" \cite{Reid-Drummond}. The situation is
similar to that of the laser above threshold, for which a correct analysis
is performed using intensity and phase variables and $not$ linearizing in
the diffusing (phase) variable. In Ref. \cite{Petrosyan} an exact
steady-state Wigner function is calculated in the single transverse-mode
NDOPO by adiabatically eliminating the pump. The phase diffusion in a mode
is evident in the radial symmetry of this distribution.  

The single
transverse-mode NDOPO is equivalent to a three mode model, describing the
extended DOPO near threshold \cite{OPOabove} through the relevant
fields $\alpha_1(\pm k_c)$and $\alpha_0(0)$ \cite{st-3modes}. The
stationary signal is \begin{eqnarray}
\alpha_1(x,t)&=&\alpha_1(k_c)e^{ik_cx}+\alpha_1(-k_c)e^{-ik_cx} \nonumber
\\ &=&2|\alpha_1(k_c)|e^{i{\theta_++\theta_-\over 2}}
\cos\left({\theta_+-\theta_- \over 2}\right), \end{eqnarray} with
$\alpha_1(\pm k_c)=|\alpha_1(\pm k_c)|e^{i\theta_\pm}$. We recognize the
effect of the  phases $\theta_\pm$ in the near field. The sum of these
 phases fixes the
global phase of the signal, locked to the pump, while the
arbitrary phase difference fixes the spatial position of the stripe
pattern. In continuous systems, where all modes are taken into account,
the diffusion of the phase difference can be interpreted as the action of the
Goldstone mode \cite{zambrini-kerr}, that is neutrally stable, 
giving a
continuous translation of the  pattern.  This is particularly evident in 
Fig. \ref{fig:1} at threshold ($E=1$).  

In a linear treatment below
threshold, the fluctuations of $\langle\theta_+ + \theta_-\rangle$ 
have  zero average and are damped
(with $\Delta_0=0$). Increasing the pump we always observe small
fluctuations, but the average changes. Fig. \ref{fig:6}b  shows that the
mean value $\langle\theta_+ + \theta_-\rangle$ 
 increases from its zero value with the distance above threshold.
This is a non-linear effect due to the feedback of the signal on the
pump.  Above threshold the average of the non-linear term in the pump
equation is not zero, so the pump is no longer real and this induces a
phase rotation in the signal.  Therefore it should be expected 
that the strongest quadrature
squeezing will be eventually
found for a local oscillator phase that depends on this
phase rotation. Here, however, we will restrict our attention to the
non-classical  features associated with intensity correlations. In fact
this corresponds to a measure of quadrature squeezing as they
result from the interference of a local oscillator given by the mean
signal field with the squeezed  fluctuations of the same mode.
 
\subsection{Intensity correlations in stripe patterns} 
\label{stripe} 

The stripe pattern formed above threshold is due to the interference of 
signal beams with opposite critical wavevectors.
Momentum conservation leads to the entanglement 
between these signal  beams
 \cite{graham,OPOabove,twin}.  This  gives
non-\-clas\-si\-cal intensity correlations characterized by  ${\mathcal
V}=-0.5$.  

\begin{figure}
\resizebox{0.99\columnwidth}{!}{\includegraphics{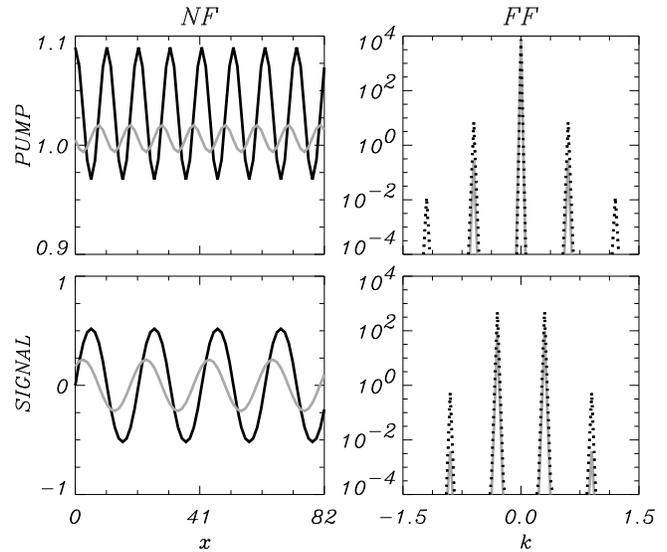}}
\caption{Snapshot of the real part of the near fields (NF) and intensity
(log scale) of the  far fields (FF) for both the pump and the signal. 
The grey plots are obtained for a pump
value $E=1.02$ and the black ones (in continuous or dotted lines)
for $E=1.1$.} \label{fig:7}
\end{figure}

Increasing the pump intensity we observe excitation of 
harmonics of the critical wavenumber
(compare far field (FF) in Fig. \ref{fig:1} for $E= 1$
and for $E= 1.1$). In Fig. \ref{fig:7} we show the real part of the near
field (NF) pattern in the pump  and in the signal 
and the corresponding FF intensities  for pumps $E=
1.02$  and $E= 1.1$. We observe that the odd harmonics are
excited in the signal and the even ones 
 in the pump mode, with an exponential decay of energy at
higher wavevector modes.  The presence of a multimode interaction means
that the momentum conservation  no longer constrains the intensities of
the twin beams, as it does below and at threshold. However, as shown in
Ref. \cite{twin}, we do observe in this regime the symmetry $\langle
N_1(k)\rangle =\langle N_1(-k)\rangle$ in intensity averages and quantum
correlations  between the critical modes survive, as shown in Fig.
\ref{fig:5} for  $1<E<1.1$. The secondary process of up-conversion of
pairs of signal photons $+k_c$ (or equivalently $-k_c$) to form pump
photons  $+2k_c$ ( $-2k_c$) does not seem to destroy the quantum
correlations between the signal ``twin" photons ($+k_c$ and $-k_c$). In
principle this process gives an incoherent depletion of the ``twin" beams,
but probably due to the smallness of this secondary effect, the
quantum correlations associated with the fundamental process survives.  

\begin{figure}
\resizebox{0.85\columnwidth}{!}{\includegraphics{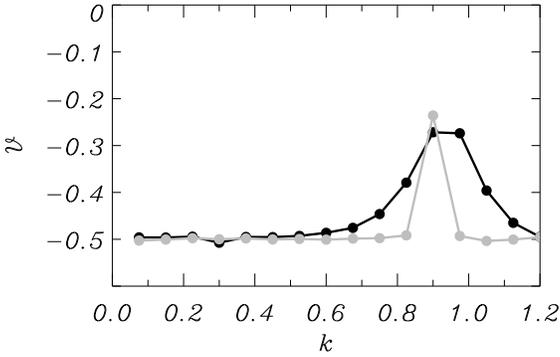}}
\caption{Twin beams correlations ${\mathcal V(k)}$ above threshold, for
$E=1.02$ (grey) and for $E=1.1$ (black).} \label{fig:8} \end{figure}

The spatial
spectrum of the intensity variance ${\mathcal V}$ is plotted in Fig.
\ref{fig:8}.  We observe that at $2\%$ above
threshold the spectrum is similar to the spectrum below threshold (compare
with  Fig. \ref{fig:4}).  
A reduction of the squeezing is observed
(peak at $k=3k_c=0.9$ in Fig. \ref{fig:8}), however, corresponding to the
appearance of the third harmonic. Increasing the pump to $10\%$ above
threshold we observe an enhancement of the spectral bandwidth in which
this reduction of squeezing appears. The  third harmonic 
in the signal is involved in
at least two important processes: the down-conversion of the homogeneous
pump into twin photons $+3k_c$ and $-3k_c$ and the secondary process of
down-con\-ver\-sion of the second harmonics  $\pm 2k_c$ of the
pump into opposite
signal photons  $\pm 3k_c$ and $\mp k_c$. We note that as the Hamiltonian
operator is Hermitian, the opposite (up-conversion) processes are also
allowed. The observed reduction of squeezing can be interpreted as a
signature of the mentioned secondary process, in which the pairs signal $+
3k_c$ and $-3k_c$ photons are incoherently (not simultaneously) generated
and destroyed. In other words, signal modes are depleted
 independently, taking
part in different cascading processes and generating harmonics. 
The entanglement should be preserved in
opposite signal modes for which the fundamental down-conversion process
prevails.  
We have confirmed that opposite
spatial modes in the pump field do not show quantum correlations.

We note that the  variances ${\mathcal V}$ obtained  with the
Q-repre\-sentation --after reordering-- are in good agreement with the
corresponding quantities calculated with the time dependent parametric
approximation in the Wigner representation in Ref. \cite{twin}.

\subsection{Intensity correlations in spatially disordered
structures} 
\label{disordered}  

\begin{figure*}
\hspace*{1.6cm}$E=0.999$\hspace{3.1cm}$E=1$\hspace{3.6cm}$E=1.1$\hspace{3.3cm}$E=1.5$\\
\resizebox{2\columnwidth}{!}{ \includegraphics{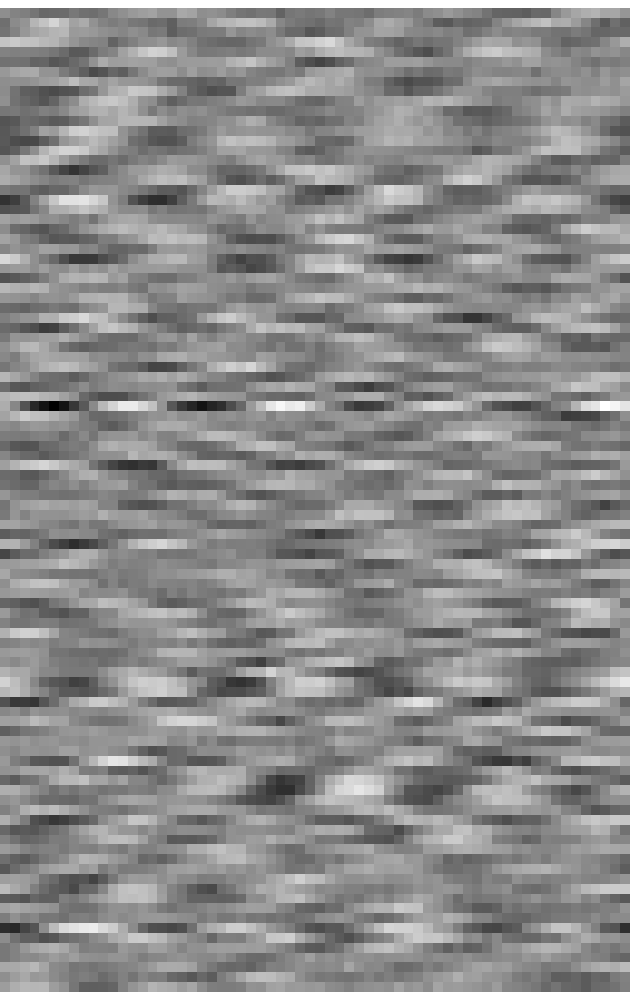} 
\includegraphics{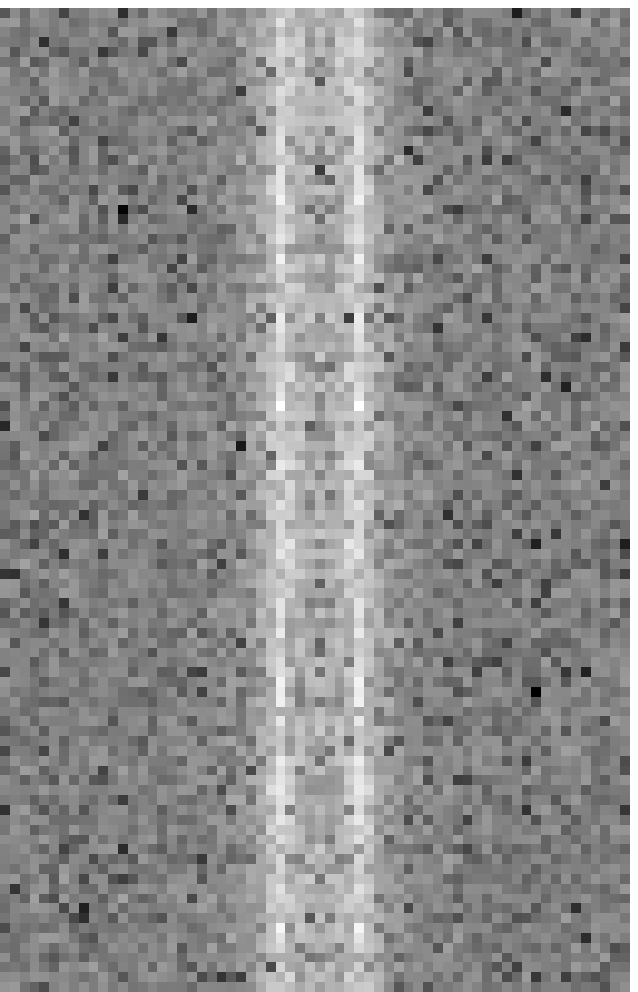} \hspace{.5cm}\includegraphics{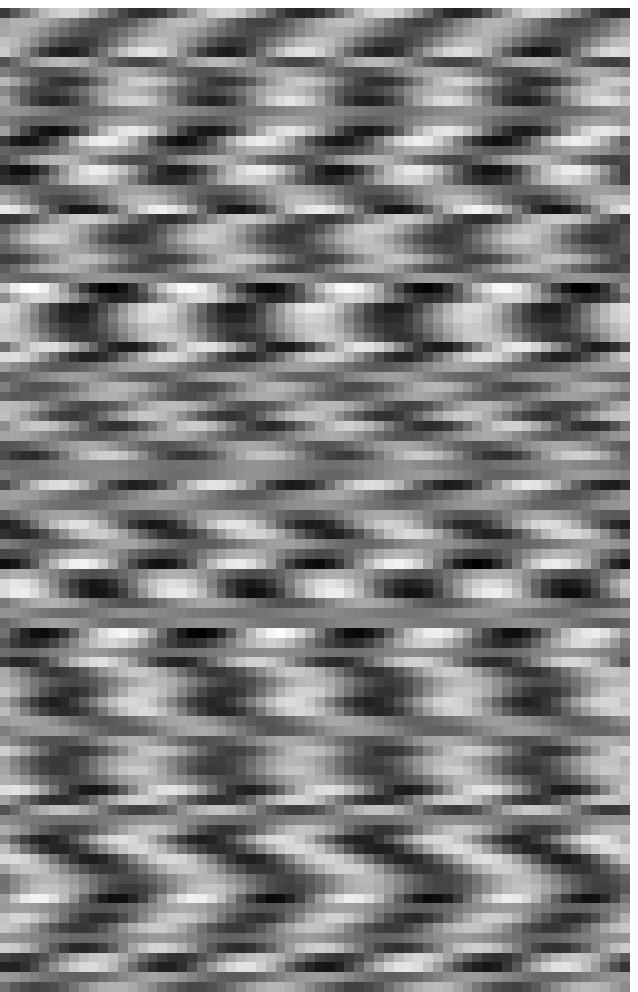}
\includegraphics{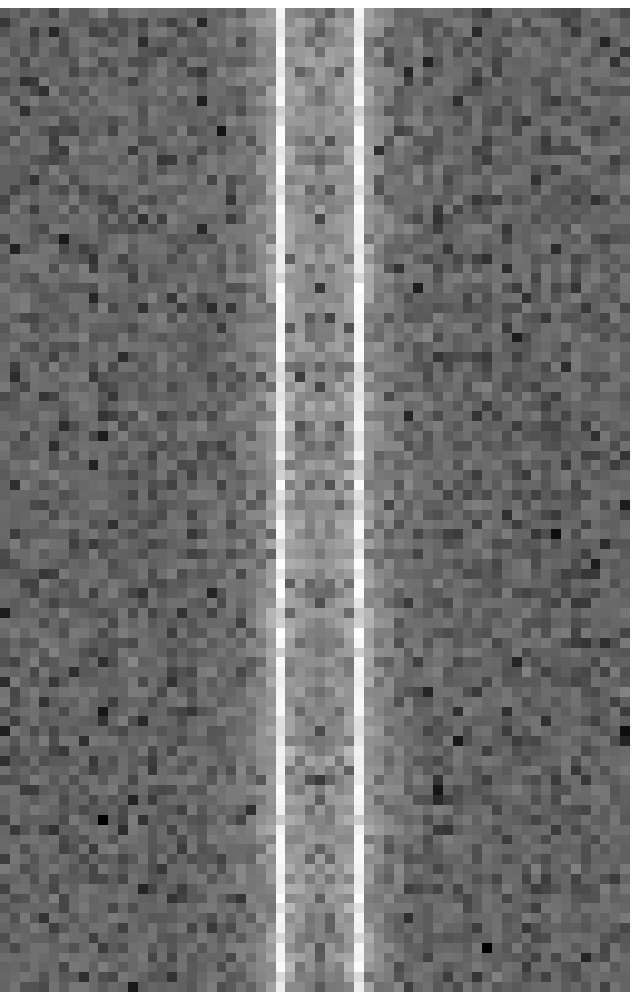} \hspace{.5cm}\includegraphics{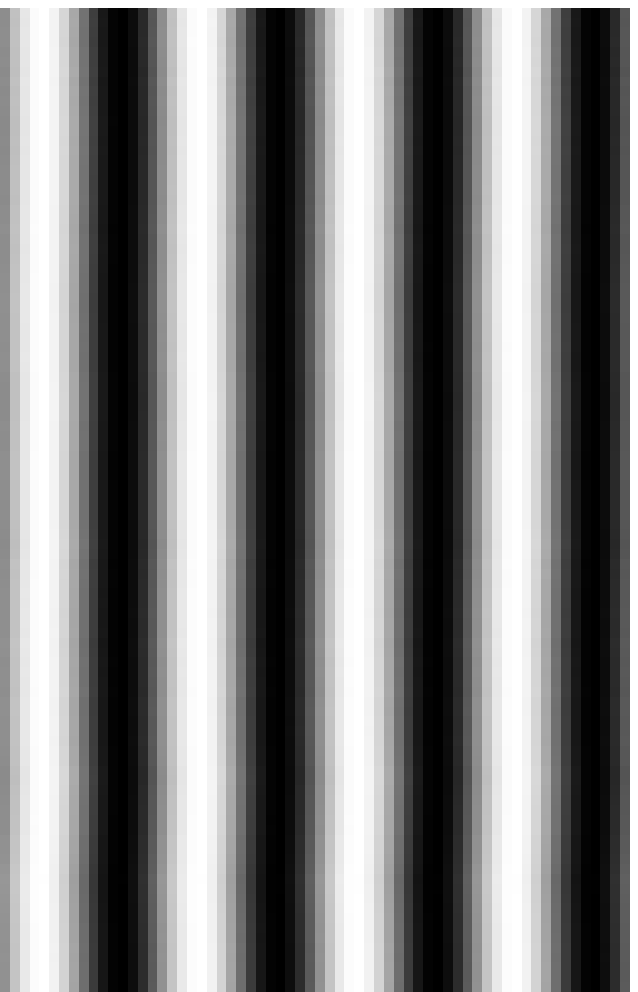}
\includegraphics{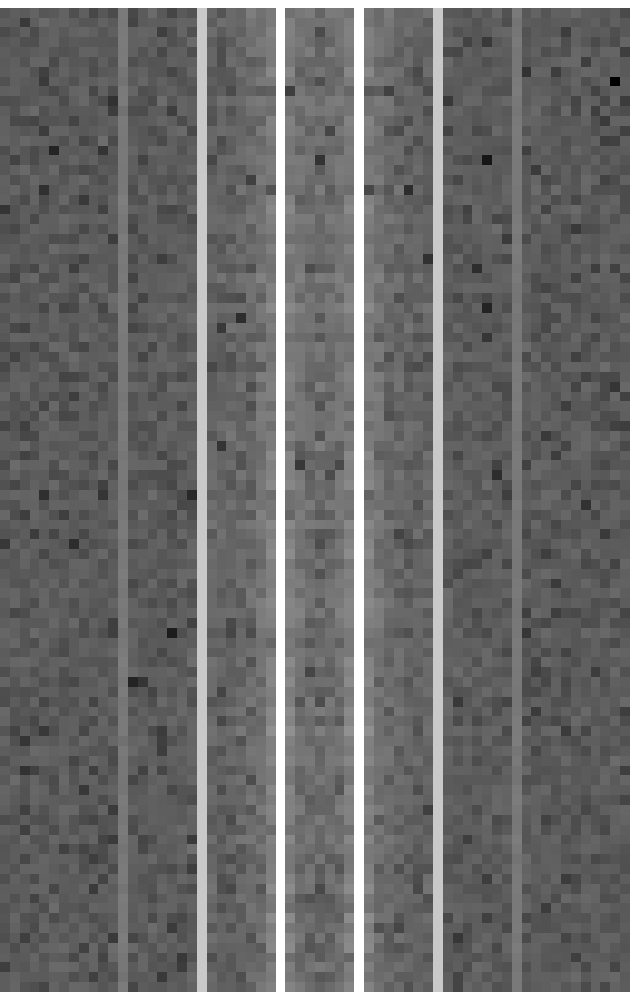} \hspace{.5cm}\includegraphics{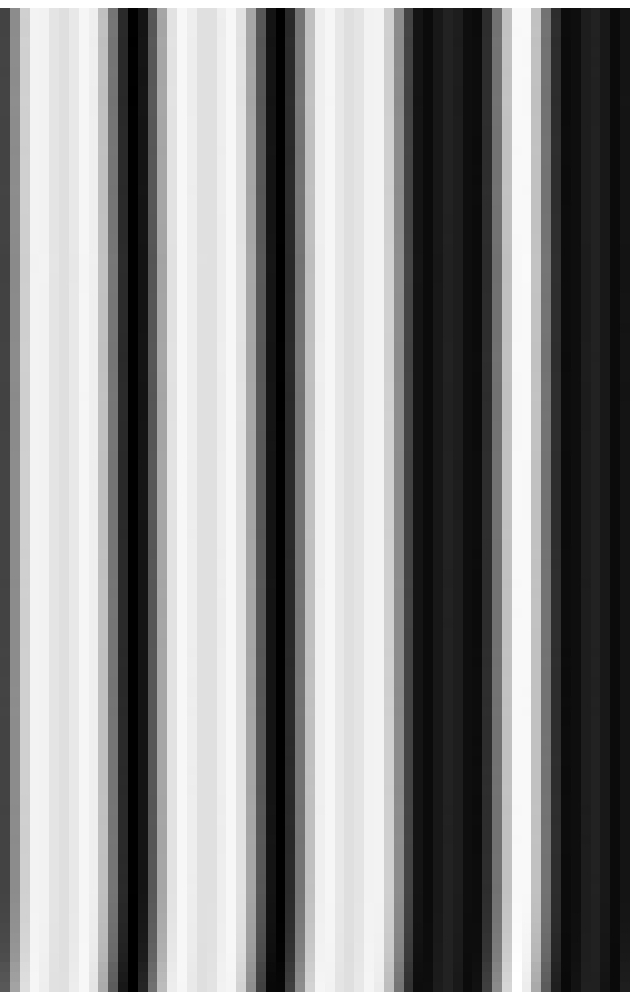}
\includegraphics{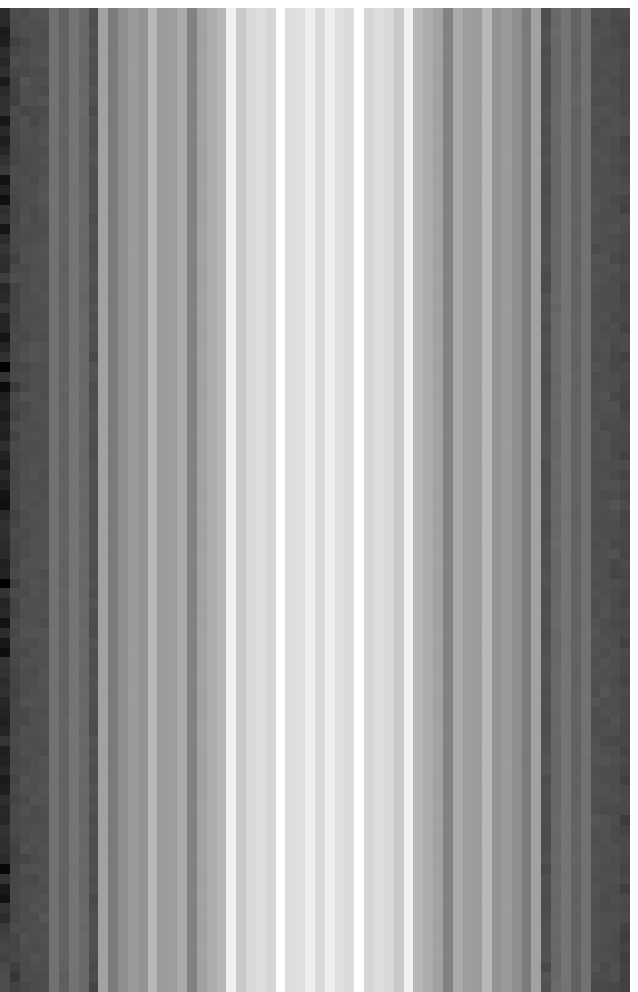}}
\\\hspace*{1cm}NF\hspace{1.6cm}FF\hspace{1.8cm}NF\hspace{1.7cm}FF\hspace{1.8cm}NF
\hspace{1.7cm}FF\hspace{1.8cm}NF\hspace{1.7cm}FF
\caption{Spatiotemporal evolution of the real part of the near field (NF)
and of the intensity of the far field (FF) (in log scale), for
different values of the pump $E=0.999,1,1.1,1.5$. 
The FF intensity is defined as $|\alpha_i( k)|^2$, 
where $\alpha_i( k)$ is the Fourier transform of the near
field  $\alpha_i( x)$.
The horizontal
coordinate is the transversal position ($x$ in NF and $k$ in FF) describes
by $64$ points, and the
vertical one is the time interval $10^7$ (in $\gamma$ units), using a discretization
time step of $\Delta t=0.01$. The initial condition for the signal is
$\alpha_1(x,0)=10^{-5}(\epsilon(x)+10\sin(k_cx))$ with $\epsilon(x)$
Gaussian random numbers of variance one.} \label{fig:1} \end{figure*}

\begin{figure}
\resizebox{0.99\columnwidth}{!}{\includegraphics{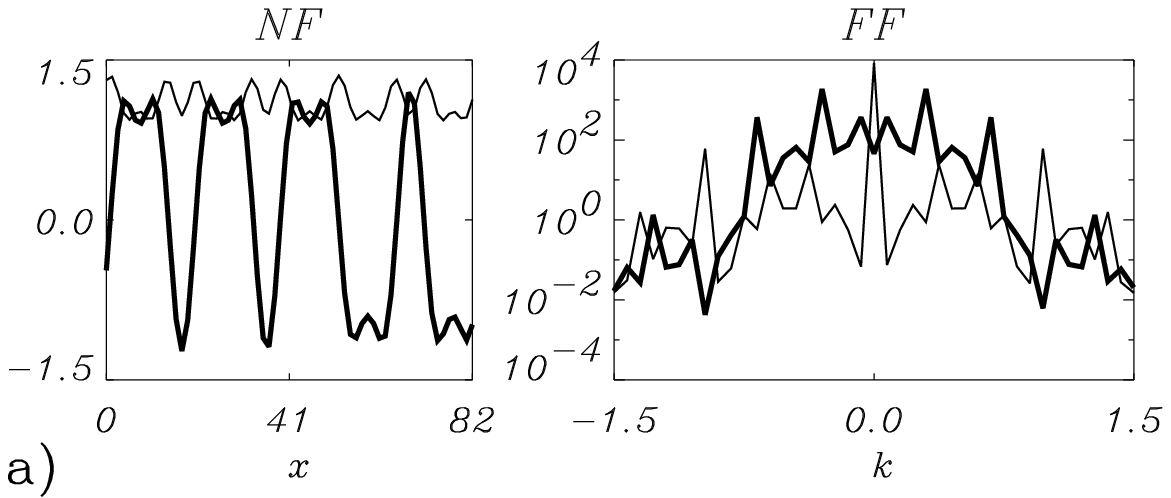}}
\resizebox{0.99\columnwidth}{!}{\includegraphics{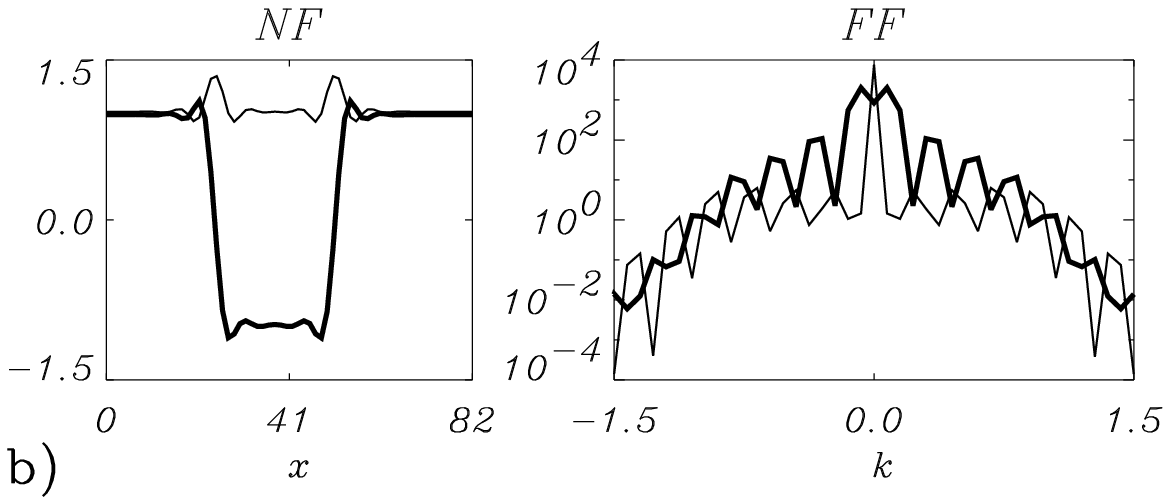}}
\caption{Snapshots of the real part of  the NF  and intensity
(log scale) of the  FF for the pump (thin line)
and the signal (thick line),
 for $E=1.5$, starting from a rolls
pattern (a) and from a step function with values $-1$ and $+1$(b).}
 \label{fig:9} 
\end{figure}

With increasing pump intensity, a
transition from a modulated pattern to homogeneous solutions takes place
 \cite{subcritical}. 
Due to the bistability in this
regime, different homogeneous solutions can be selected in separated
spatial domains (see  Fig. \ref{fig:1} for $E=1.5$). 
All solutions presented in Fig.
\ref{fig:1} are obtained starting from a modulated initial condition at
$k_c$.  Therefore we are $stimulating$, with
a particular initial configuration, the shape of the final structure. In
Fig. \ref{fig:9}a  we show the stationary configuration for
$E=1.5$, in which the most excited mode is $k_c$.  
The disordered character of the structure gives rise to a broad
spectrum, in which the other dominant modes are not harmonics of $k_c$. 
The most
intense signal modes combine to form pump modes: for example in Fig.
\ref{fig:9}a the signal modes $k_c=4\Delta k=0.3$ and 
$9\Delta k=0.675$ give the
pump mode $13\Delta k=0.975$ \cite{Dk}. The stationary disordered structure
shows the same reflection symmetry $\langle N_1(k)\rangle =\langle
N_1(-k)\rangle$, of the stripe pattern considered  in  Sect.
\ref{stripe}.  

\begin{figure}
\resizebox{0.9\columnwidth}{!}{\includegraphics{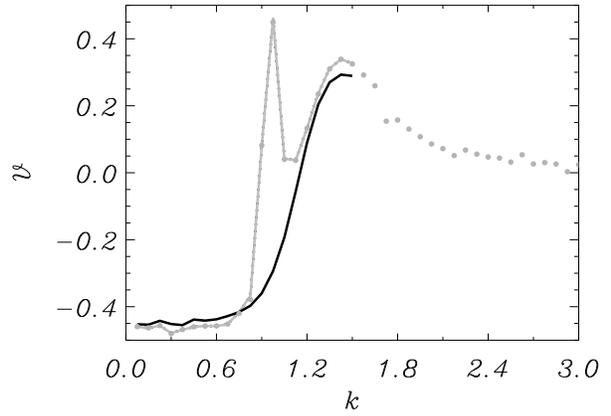}}
\caption{Twin beams correlations ${\mathcal V(k)}$ for pump $E=1.5$. The
grey (black) line is the spectrum ${\mathcal V(k)}$ for the
pattern shown in Fig. \ref{fig:9}a  (Fig. \ref{fig:9}b ). The grey dotted
line is obtained considering a system with the same size ($L=4\lambda_c$)
but with a finer discretization (128 instead of 64 points), starting from
a stripe configuration of critical wave-length.
In this way we can see the asymptotic behavior
of the spectrum for large wavevectors. For small wave-vectors the results
are in good agreement with the simulation using 64 points.} \label{fig:10}
\end{figure}

Studying the properties of the quantum fluctuations in this
regime, we observe non-classical correlations between the two signal
``twin" beams with critical wave-vectors, also in these spatially disordered
structures (see Fig. \ref{fig:5} for $1.2 \leq E \leq 1.5$). This result
does not depend on momentum conservation or on the presence of a regular
pattern. 
We have also considered the entanglement properties of modes different 
from the critical one.
In Fig.
\ref{fig:10} we show the spatial spectrum of the variance
${\mathcal V}$.  As in the case of a regular stripe pattern, analyzed in
the previous section, we continue to
find quantum correlated twin beams. However there are some interesting 
differences.
The peak in  ${\mathcal V}$ now corresponds to a
strongly depleted ($low$ mean intensity) signal mode for 
$k=0.975$.  This
contrasts with the previous case, where $k=3k_c$ was an excited mode. 
The most interesting feature is the appearance of a bandwidth of
``twin" beams, where the signal field is intense;  the correlations
become classical  for big wave-vectors ($k\gtrsim 1$), 
where the signal is depleted more than the pump
 field, reaching asymptotically the
level of coherent states (see dotted grey line in Fig. \ref{fig:10}). In
conclusion, the ``twin"-beams quantum correlations persist in 
disordered structures.  This signature of the fundamental
down-conversion process is preserved throughout 
the region of intense signal modes.  

The demanding question is how the  spectrum of the variance
${\mathcal V}$ is influenced by the shape of the selected spatial
structure. In this regime no special character is associated 
with the critical wave-length periodicity.
Therefore  we consider a stationary state
of two domains, obtained from an initial step condition (Fig.
\ref{fig:9}b). Also in this case, with a very different stationary
state, we observe non-classical  intensity correlations in the bandwidth
$0<k\lesssim 1$, as shown in  Fig. \ref{fig:10} (black line). 
In this case no peaks appear, suggesting that the
presence of modes with reduced squeezing (as the peak for $k= 0.975$
observed in the previous structure) depends on the selected spatial
structure. 

\begin{figure}
\resizebox{0.7\columnwidth}{!}{\includegraphics{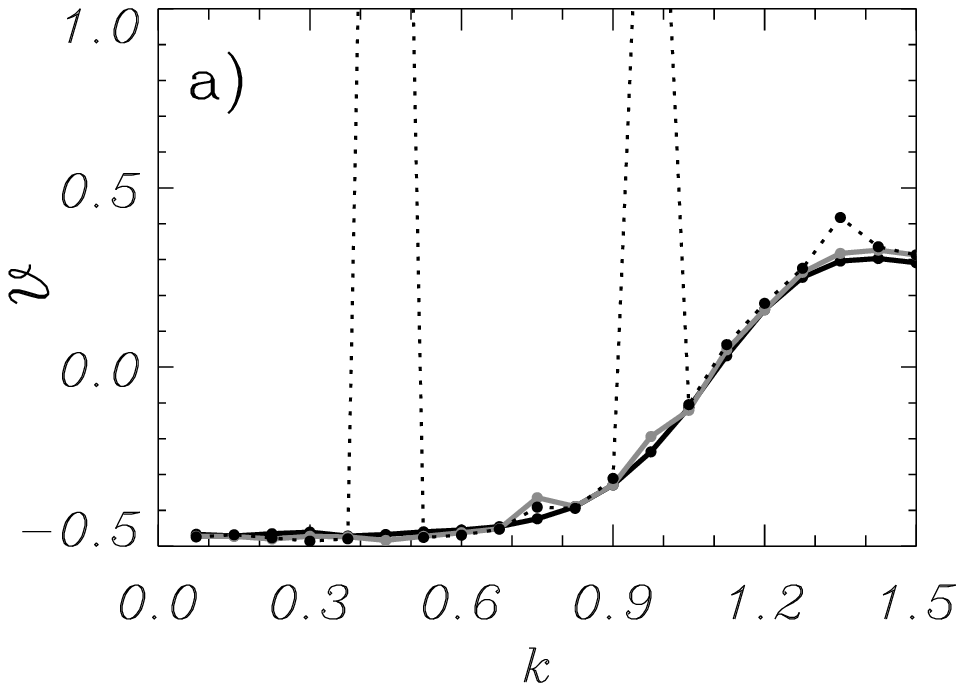}}
\resizebox{0.99\columnwidth}{!}{\includegraphics{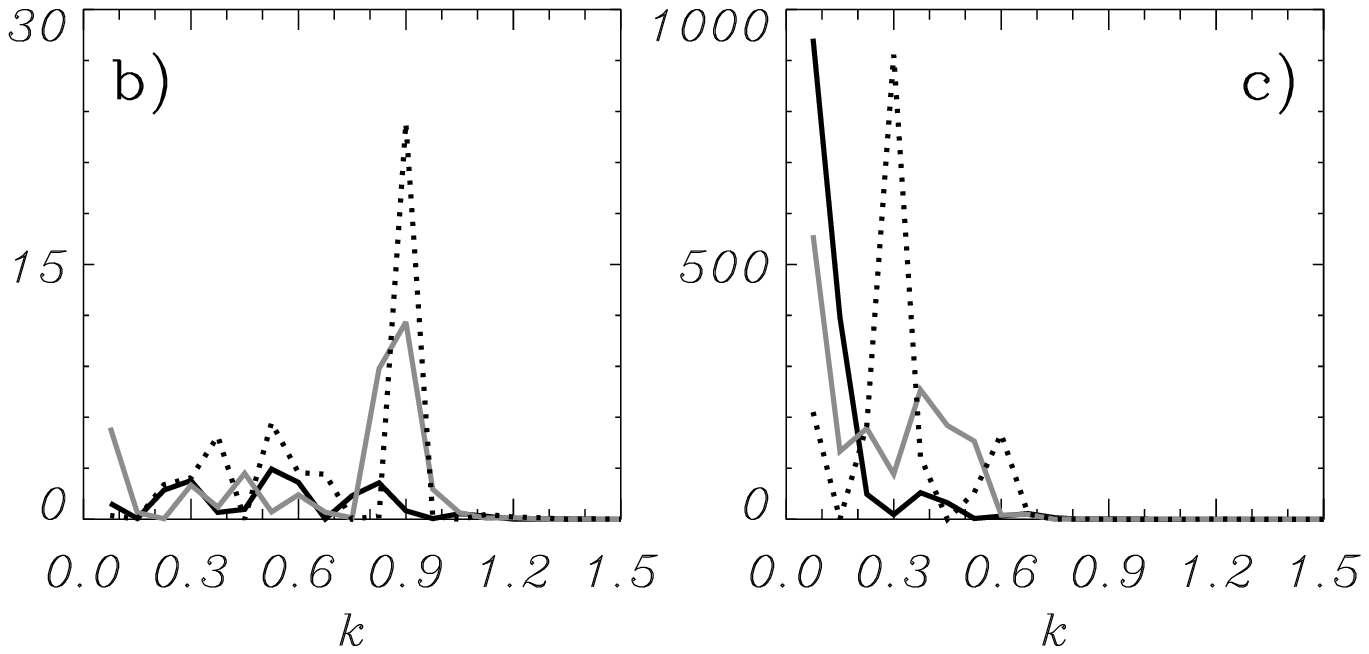}}
\caption{a) Twin beams correlations
 ${\mathcal V}(k)$  for pump $E=1.3$,
corresponding to different spatial structures, obtained starting from a
step function (black line), from noise (grey line) and from a stripe pattern
with critical wave-length (dotted line). Mean FF intensity in the
pump ($<N_0>$) (b) and in the signal ($<N_1>$) fields (c), 
for the same three spatial configurations.} \label{fig:11} 
\end{figure}

The key point is the relative importance of the fundamental
coherent process of twin photon down-conversion, and other incoherent
cascading processes, which depends on the spatial configuration chosen by the
system. To give more evidence of this statement, we study the variance
${\mathcal V}$ in three different spatial configurations, obtained for
$E=1.3$, starting the simulations from noise, from rolls (as in Fig.
\ref{fig:9}a) and from a step function (as in Fig. \ref{fig:9}b).  In Fig.
\ref{fig:11} we show the spectral variances (a) and the mean FF
intensity (b). The domain configurations obtained staring
from noise  and from a step function give
overlapping smooth variances, while two peaks appear when starting from
rolls . Also in this case the peaks appear associated with
strongly depleted signal modes with low intensities, 
and can be in different
positions depending on the selected spatial structure. The bandwidth of
non-classical variance ${\mathcal V}$ seems 
to be a general feature, almost independent of the
structure selected. It corresponds to the FF region
of intense tilted signal beams.

\section{Conclusions}  
\label{conclusions}
Non-linear optical systems present a wealth of
physical phenomena including self-organizing spatial patterns and quantum
correlations. Commonly employed methods to study these phenomena
include the use of the positive P representation and the Wigner representation
with third order derivatives neglected.  Here we have investigated the use
of the Q representation for studying quantum correlations in the
DOPO at and above threshold.   Positive diffusion is not guaranteed with the Q
representation and this can  lead to problems with divergent
trajectories.  For the DOPO, however, only positive diffusion occurs
unless the fluctuations are strong enough to push the pump field up to
twice the threshold value. This never occurred in our simulations and we
have not attempted  to calculate the effects of such highly unlikely
trajectories on our ensembles \cite{Aspect}.

Below threshold entanglement in quadratures and non-classical intensity
correlations are obtained, in agreement with linear analytical results.
At threshold, a pair
of quantum correlated twin beams  is generated.  These beams have
wavevectors $\pm k_c$, corresponding to the critical wavelength.  The
quantum correlations are a natural consequence of the fundamental
microscopic process in which a single pump photon is converted into a pair
of signal photons. As we move further above threshold the twin beams can
recombine to generate new pump photons.  The combination of a $+k_c$ and a
$-k_c$ photon regenerates one of the original pump photons.  The
combination of a pair of $+k_c$ photons or a pair of $-k_c$ photons,
however, is a new process and introduces higher harmonics in the pump and
thence  in the the signal.  Such processes can also degrade, but not
completely suppress, the quantum correlations in some signal
modes. This is a signature of the incoherent depletion
of those modes. Yet further
above threshold we enter a regime of spatially disordered  structures. 
Remarkably, quantum correlations persist even in this regime,
in the bandwidth of intense signal modes, where they
take a form that depends on the spatial pattern that is generated. 

We acknowledge financial support from the European Commission
project QUANTIM (IST-2000-26019).
RZ, PC and MSM acknowledge financial support from  the Spanish
MCyT projects PB97-0141-C02-02 and BFM 2000-1108.
SMB thanks the Royal Society of Edinburgh and
the Scottish Executive Education and Lifelong Learning Department
for financial support.

\newpage


\begin{thebibliography}{}
\bibitem{notetb} Note that even far below threshold, where the critical
mode does not play an important  role, any two opposite modes $\vec{k}$
and $-\vec{k}$ are entangled.
\bibitem{stevenote} Here $\Delta_i$ and $a_i$
are cavity detunings and  diffraction parameters for the pump and signal
expressed in units of  the associated decay constants $\gamma_i$
\bibitem{doubling} The phase space doubling  introduced with the $P_+$
implies a much larger computational effort in the simulation of the
Langevin equations. This is a critical problem with spatial degrees of
freedom and justifies the search for alternative representations and
methods.
\bibitem{obs} When the condition for a positive diffusion
 (Eq. \ref{d.positive.sc}) is fulfilled,
the arguments of the square roots in Eq. (\ref{x1}) are all positive.
\bibitem{shot-noise} The shot noise level $\mathcal{N}_X$ 
for the variance of  the quadrature superpositions  is associated to the
commutator of creation and destruction operators. 
A detailed evaluation of this quantity can be found in Ref.
\cite{zambrini}.
\bibitem{st-3modes} The deterministic steady
state above threshold gives  $|\alpha_1(k_c)|=|\alpha_1(-k_c)|$ and,
with our choice of parameters, the phase sum $\theta_++\theta_-=0$.
\bibitem{subcritical} The transition depends effectively 
on the noise level. For example, for pump $E=1.1$,
 for which we have analyzed in
Sect. \ref{stripe} fluctuations in stripe patterns, a high level of noise 
($g/\gamma=10^{-2}$) leads to a disordered structure.
\bibitem{Dk} The discretization $\Delta
k$ depends on the size of the system: $\Delta
k=2\pi/L=2\pi/4\lambda_c$. With our choice of parameters $\Delta
k=0.075$.
\end{thebibliography}
\end{document}